\documentclass[aps,prl,preprint,superscriptaddress,letterpaper,floats,floatfix,letterpaper,floatfix]{revtex4-1}

\usepackage{amsmath}
\usepackage[pdftex]{graphicx}
\usepackage[thinspace]{SIunits}
\usepackage{color}
\usepackage{braket}
\usepackage{graphicx}

\newcommand{\bB}{\mathbf{B}}

\newcommand{\bb}{\mathbf b}

\newcommand{\bs}{\mathbf{s}}

\newcommand{\bn}{\mathbf n}

\newcommand{\bS}{\mathbf{S}}
\newcommand{\be}{\begin{eqnarray}}
\newcommand{\ee}{\end{eqnarray}}
\newcommand{\la}{\langle}
\newcommand{\ra}{\rangle}

\newcommand{\da}{\downarrow}
\newcommand{\ua}{\uparrow}

\begin{document}

\title{Quantum Fingerprints in Higher Order Correlators of a Spin Qubit}

\author{Alexander Bechtold}
 \affiliation{Walter Schottky Institut and Physik Department, Technische Universit\"at M\"unchen, 85748 Garching, Germany}
\author{Fuxiang Li}
 \affiliation{Center for Nonlinear Studies, Los Alamos National Laboratory,  Los Alamos, New Mexico 87545 USA}
 \affiliation{Theoretical Division, Los Alamos National Laboratory, Los Alamos, New Mexico 87545, USA}
\author{Kai M\"uller}
 \affiliation{Walter Schottky Institut and Physik Department, Technische Universit\"at M\"unchen, 85748 Garching, Germany}
 \affiliation{E. L. Ginzton Laboratory, Stanford University, Stanford, California 94305, USA}
\author{Tobias Simmet}
 \affiliation{Walter Schottky Institut and Physik Department, Technische Universit\"at M\"unchen, 85748 Garching, Germany}
\author{Per-Lennart Ardelt}
 \affiliation{Walter Schottky Institut and Physik Department, Technische Universit\"at M\"unchen, 85748 Garching, Germany}
\author{Jonathan J. Finley}
%%% \email{jonathan.finley@wsi.tum.de}
 \affiliation{Walter Schottky Institut and Physik Department, Technische Universit\"at M\"unchen, 85748 Garching, Germany}
\author{Nikolai A. Sinitsyn}
 \affiliation{Theoretical Division, Los Alamos National Laboratory, Los Alamos, New Mexico 87545, USA}

\maketitle

\textbf{
The spin of an electron  in a semiconductor quantum dot represents a natural nanoscale solid state qubit \cite{Loss1998, Hanson2008, Edamatsu2008, Ladd2010}. Coupling to  nuclear spins leads to decoherence that limits the number of allowed quantum logic operations for this qubit. Traditional approach to characterize decoherence is to explore spin relaxation and the spin echo, which are equivalent to  the studies of the spin's 2$^\text{nd}$ order  time-correlator at various external conditions.  Here we develop an alternative technique by showing that considerable information about dynamics can be obtained from direct measurements of higher than the 2$^\text{nd}$ order correlators, which to date have been hindered in semiconductor quantum dots. We show that such correlators are sensitive to pure quantum effects that cannot be explained within the classical framework, and which allow direct determination of ensemble  and quantum dephasing times, $T_2^*$ and $T_2$, with only repeated projective measurements without the need for coherent spin control. Our method enables studies of pure quantum effects, including tests of the Leggett-Garg type inequalities that rule out local hidden variable interpretation of the spin qubit dynamics.}

Electronic spins in InGaAs quantum dots (QDs) have shown long spin life-times  $T_1$, up to milliseconds \cite{Heiss2009, Khaetskii2002}, and intrinsic dephasing times $T_2$ in the microsecond range, as measured by spin echo experiments in strong magnetic fields \cite{Alex, Press2010, DeGreve2011, Bluhm2010}. Moreover, optical polarization of a nuclear spin bath can extend the central spin lifetime up to an hour \cite{hour-long, Zhong2015}. This indicates a considerable potential of QD-qubits for quantum information processing. However,  spin echo is a classical effect in the sense that it can be fully explained in terms of a classical measurement and the behavior of classical spins changing the direction of their precession under the action of properly applied control pulses \cite{Economou2007}. Thus, available data for central spin relaxation, spin echo, and  spin fluctuations  \cite{ingaas3,ingaas4,ingaas5,ingaas6} could be well explained, so far, purely within the semiclassical approach \cite{Merkulov2002, Al-Hassanieh2006, Zhang2006, Testelin2009, Sinitsyn2012}.
Considering that an electron confined in a quantum dot is a quasiparticle dressed by continuous virtual interactions with other solid state excitations,  a long spin relaxation time $T_1$ and spin echo life-time $T_2$ do not necessarily predicate the ability of QD spins to process quantum information at these time scales.

We propose an alternative route to characterize the quantum nature of a solid state qubit. By ``quantum" we mean  effects that cannot be explained without resorting to the quantum measurement theory. We will identify such effects by introducing a measurement technique that can determine, in principle, an arbitrary order correlator $\la \hat{Q}_{t_{n}+\ldots +t_1} \ldots \hat{Q}_{t_1} \hat{Q}_{0} \ra$, where sub-indices indicate the time moments of application of the projection operator acting on the electronic spin, defined as $\hat{Q}\equiv \ket{\downarrow} \bra{\downarrow}$, and $\la \ldots \ra$ indicates an averaging over many repeated pulse sequences. 

\begin{figure}[t]
\centering
\includegraphics[width=.6\textwidth]{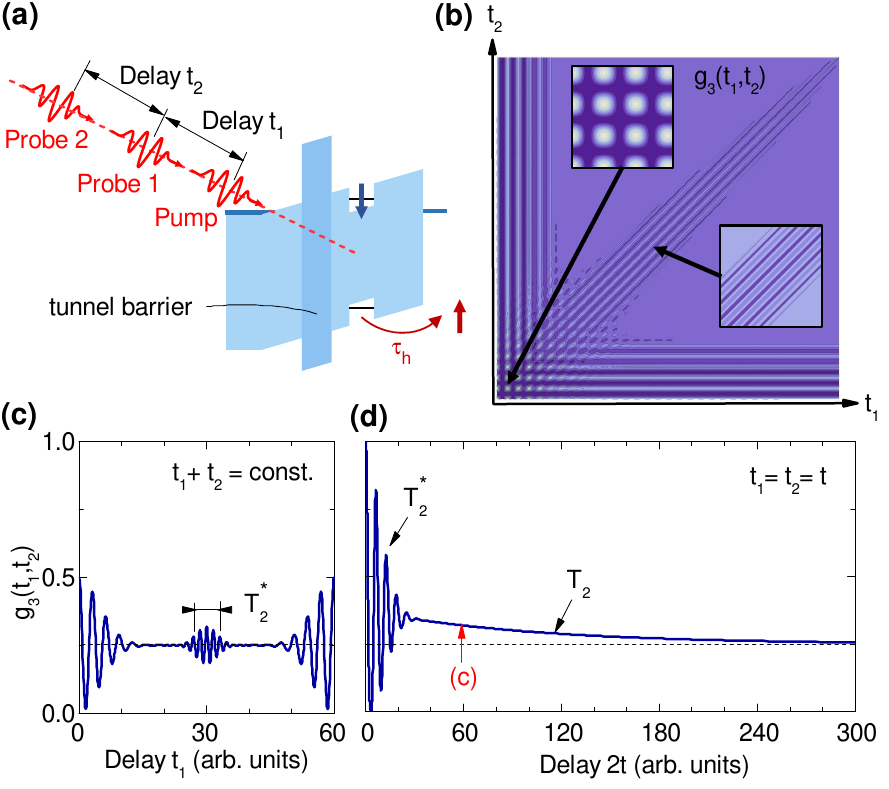}
\caption{\label{fig1}
\textbf{Experimental method and theoretical predictions of the outcome of a three pulse experiment.}
\textbf{(a)} The optical pump pulse prepares the electron in a spin down state, applications of optical probe pulses determine the electronic spin state at time moments $t_1$, $t_1+t_2$. The QDs are embedded in a voltage tunable spin memory device. \textbf{(b)} Contour plot of $g_3 (t_1,t_2)$ using Eq.~(\ref{g3-3}) with $R(t-t')=(1/T_2^*)^2+(2/T_2) \delta(t-t')$. Insets show details of $g_3$ at short and long timescales.   \textbf{(c)} Cut of $g_3$ along anti-diagonal direction, $t_1+t_2=$ const. \textbf{(d)} Cut of $g_3$ along diagonal direction, $t_1=t_2$.
}
\end{figure}

The idea of our experimental method is illustrated in Fig.~\ref{fig1}(a). An electron spin confined in a single self-assembled InGaAs QD can be prepared in the $\ket{\downarrow}$ state using a single picosecond laser pulse \cite{Heiss2009, Jovanov2011, Kroutvar2004}, as indicated with ``Pump" in the figure. This is equivalent to the nonzero outcome of the application of the projection operator $\hat{Q}_{0}$ at the initial moment in time. By taking advantage of an asymmetric tunnel barrier, the electron spin can be trapped in the QD over timescales extending up to seconds \cite{Heiss2009}. Following the electron spin initiation, we apply one or more circularly polarized laser pulses (``Probe~1", ``Probe~2", etc.) that probe the state of the spin qubit at later moments ($t_1$, $t_2$, etc.). If the electronic spin is in the state $\ket{\uparrow}$, such a probe pulse excites an additional electron-hole pair, and the QD becomes charged with two electrons ($2e$) and is, therefore, optically inactive during the whole remaining time of the experiment. Such a state corresponds, in our notation, to the zero outcome of the measurements described by the projection operator $\hat{Q}_{t}$. On the other hand, if the electron spin is in the state $\ket{\downarrow}$ before the application of the probe pulse, the Pauli principle does not allow the excitation of a second electron-hole pair such that the probe pulse becomes essentially unnoticed by the electron. At the end of the pulse sequence, we perform the measurement of the total charge in the QD (not shown in Fig.~\ref{fig1}(a)). Here, an observation of a doubly charged QD corresponds to at least one zero outcome of the measurements by operators $\hat{Q}$ applied at the instants in time of the optical pulses. Conversely, finding a singly charged QD corresponds to results $Q=1$ in all measurement pulses. We provide technical details in the Supplemental Material.

The preparation pulse sets the spin density matrix at $\hat{\rho}=  \ket{\da}\bra{\da}$, which is equivalent to the nonzero measurement outcome by the operator $\hat{Q}$ at $t=0$.  Let $\hat{G}(t)$ be the evolution matrix for the measurement probabilities with an element $G_{\alpha \beta}(t)$, $\alpha,\beta \in \{ \uparrow,\downarrow \}$, meaning the probability that after the system starts at state $\beta$, the measurement of the spin state along the $z$-axis at time $t$ afterwards would find the spin in the state $\alpha$. Then the second order correlator would be 
\be
g_2(t) =  {\rm Tr} \Big[   \hat{Q} \hat{G}(t) \hat{Q} \Big],
\label{g2}
\ee
and the third order correlator is
\be
g_3(t_1,t_2) = {\rm Tr} \Big[  \hat{Q}\hat{G}(t_2) \hat{Q} \hat{G}(t_1) \hat{Q} \Big].
\label{third2}
\ee

Between measurements, the presence of an external magnetic field leads to oscillations of the probability of observing the $Q=1$ outcome. Importantly, even if this value is observed, quantum measurement is generally destructive, i.e. it resets the density matrix to the one of the pure state,  $\ket{\da}\bra{\da}$. An exception is when the time intervals $t_1$ and $t_2$ are chosen to be commensurate with the period of the spin precession. This situation corresponds to the resonant enhancement of the correlator $g_3(t_1,t_2)$ \cite{Liu2010}.
The latter property becomes especially pronounced in $g_3$ for times $t_2=t_1 \gg T_2^*$. Consider, for example, the precession of the spin in a time-dependent magnetic field applied transverse to the measurement axis direction: 
\be
G_{\da \da}(t) =\frac{1}{2}\left( 1 +   \cos \left(\int_0^t \omega(t') \, dt' \right) \right), 
\label{prec}
\ee
where  $\omega$ is the precession frequency.
 We  assume that $\omega(t)$ has a  strong constant component due to an external field with Larmor frequency $\omega_L$, and a fluctuating component due to the dynamics of the Overhauser field with frequency $\omega_O$: $\omega=\omega_L + \omega_O(t)$. The latter has nearly Gaussian statistics: $\la \omega_O(t) \omega_O(t') \ra = R(t-t')$, with some correlation function $R(t)$ \cite{Sinitsyn2012}. 
 Substituting (\ref{prec}) into (\ref{g2})-(\ref{third2}) and averaging the result over the Overhauser field distribution we find
\be
\label{g2-3} g_2(t) = \frac{1}{2} \left(1+\cos(\omega_L t) e^{-\frac{1}{2} \int_0^t dt_1 \int_0^t dt_2\, R(t_1-t_2)}  \right), %\\ %e^{-\frac{1}{2}\Delta^2 t^2-\gamma_s t} \cos(\omega_L t),\\
\ee
\be
\label{g3-3}
\nonumber    g_3(t_1,t_2) =  \frac{1}{2} \left[ g_2(t_1)+g_2(t_2) + \frac{1}{2} g_2(t_1+t_2) \right]  - \frac{3}{8}  +
\ee
\be
+ \frac{1}{8} \cos(\omega_L(t_1-t_2)) e^{-\int \limits_0^{t_1+t_2} dt' \int \limits_0^{t_1+t_2} dt'' \frac{q(t')q(t'')}{2} R(t'-t'')}.  
\ee 
Here $q(t)=1$ for $t<t_1$ and $q(t)=-1$ for $t>t_1$. In Fig.~\ref{fig1}(b) an example of $g_3(t_1,t_2)$  is plotted using Eq.~(\ref{g3-3}), considering the case of a correlator $R(t-t')=(1/T_2^*)^2+(2/T_2) \delta(t-t')$ \cite{Liu2010}. The corresponding correlators $g_2(t)$ in Eq.~(\ref{g2-3}), and hence the term $[\ldots ]$ in (\ref{g3-3}), decay quickly during the ensemble dephasing time $T_2^*$. The inset in Fig.~\ref{fig1} shows details of $g_3$ at small and large timescales. Remarkable is the survival of the last term in Eq.~(\ref{g3-3}) along the diagonal direction, $t_1=t_2\gg T_2^*$. If $g_3$ could be expressed via the $2^\text{nd}$ order spin correlators at the equilibrium, $g_3$ would also decay quickly for $T_2\gg t_1,t_2 \gg T_2^*$ to a constant value $1/4$. However, the $3^\text{rd}$ order correlator is influenced by quantum effects (see Supplementary Section 2) that, in our case, make  the last term in Eq.~(\ref{g3-3}) immune to inhomogeneous broadening for equal time intervals between successive measurements.

\begin{figure}[t]
\centerline{\includegraphics[width=.6\textwidth]{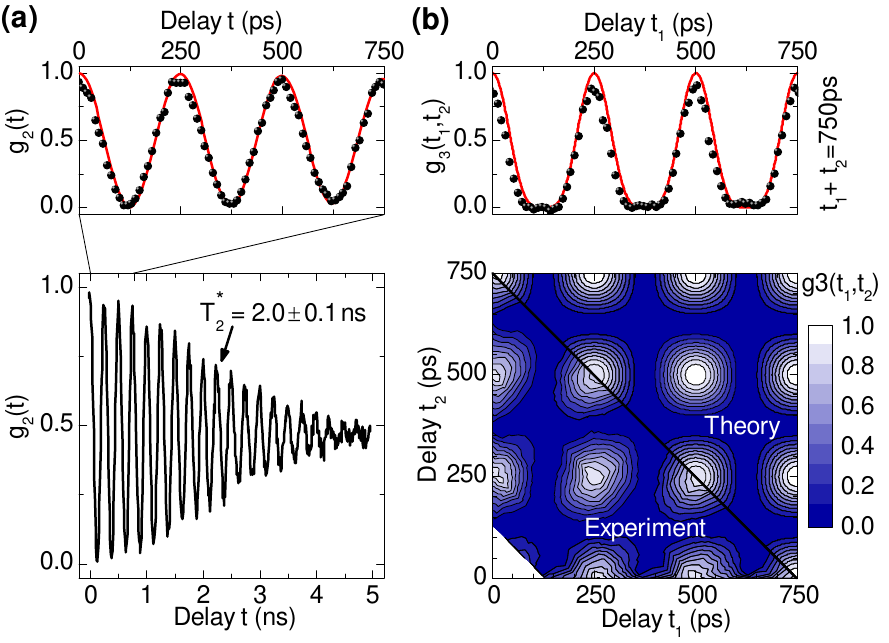}}
\vspace{0.2 cm}
\caption{\label{fig2}
\textbf{Short time behavior of $g_2$ and $g_3$ correlator at $B_x=0.5~\tesla$.}
\textbf{(a)} Experimental data of $g_2(t)$  with zoom-in over the initial picoseconds. Inhomogeneous dephasing after $2~\nano\second$ takes place owing to contributions of randomly orientated Overhauser fields. \textbf{(b)} Experimental data of $g_3(t_1,t_2)$. The upper part shows a line cut of the contour plot along the anti-diagonal direction, keeping the total time fixed to $t_1+t_2=750~\pico\second$. Comparison with theoretical predictions using Eq.~(\ref{g3-3}) (red line and upper part of contour plot).
}
\end{figure}

Along the line $t_1=t_2 \equiv t$, the correlator $g_3$ decays as $\sim e^{-2 t/T_2}$, which can be used to determine the intrinsic spin relaxation time. In fact, one can recognize the exponent in the last term in Eq.~(\ref{g3-3}), at $t_1=t_2$, as the expression that describes the spin echo amplitude in the noisy field model \cite{Alex}.  For a better visibility we show in Figs.~\ref{fig1}(c) and (d) cuts of the $g_3$ contour plot along, respectively, the anti-diagonal ($t_1+t_2=\text{const.}$) and diagonal ($t_1=t_2$) directions. Along the direction $t_1+t_2=\text{const.}$, the third order correlator oscillates with an envelope given by a Gaussian function with a width corresponding to the inhomogeneous dephasing time $T_2^*$. Along $t_1=t_2$, the oscillation amplitude first decays quickly within the inhomogeneous dephasing time $T_2^*$, then it decays slowly at the timescales of $T_2$.
 
In order to test the theoretical predictions for the correlators $g_2(t)$ and $g_3(t_1,t_2)$ experimentally, we use the spin storage device \cite{Heiss2009,Alex} and the experimental method as introduced above. Within short timescales for which $t_{1,2}$ are on the order of nanoseconds, the results of measuring $g_2$ and $g_3$ correlators are presented in Fig.~\ref{fig2}(a) and (b), respectively. As can be seen in Fig.~\ref{fig2}(a), the amplitude of $g_2(t)$ oscillates with the Larmor frequency ($|g_e|=0.55$), since an in-plane magnetic field of $B_x=0.5~\tesla$ is applied. Within the initial $2.0~\nano\second$ the amplitude of $g_2$ quickly decays with a Gaussian envelope as $\sim e^{-\frac{1}{2}(t/T_2^*)^2}$ owing to contributions of randomly oriented Overhauser fields \cite{Alex}. The red line shows the application of Eq.~(\ref{g2-3}), demonstrating the high fidelity of our spin initialization and readout methods, a necessary pre-requisite for conducting higher order correlation measurements. In contrast to the sinusoidal behavior of $g_2$, the correlator $g_3$ in a three pulse experiment shows a pattern that is comparable to $g_2(t_1) g_2(t_2)$ at such short timescales, as depicted in Fig.~\ref{fig2}(b), and agrees very well with the theoretical predictions of Eq.~(\ref{g3-3}) (red line and contour plot in Fig.~\ref{fig2}(b)).

\begin{figure*}[t]
%\vspace*{.05in}
\centerline{\includegraphics[width=1\textwidth]{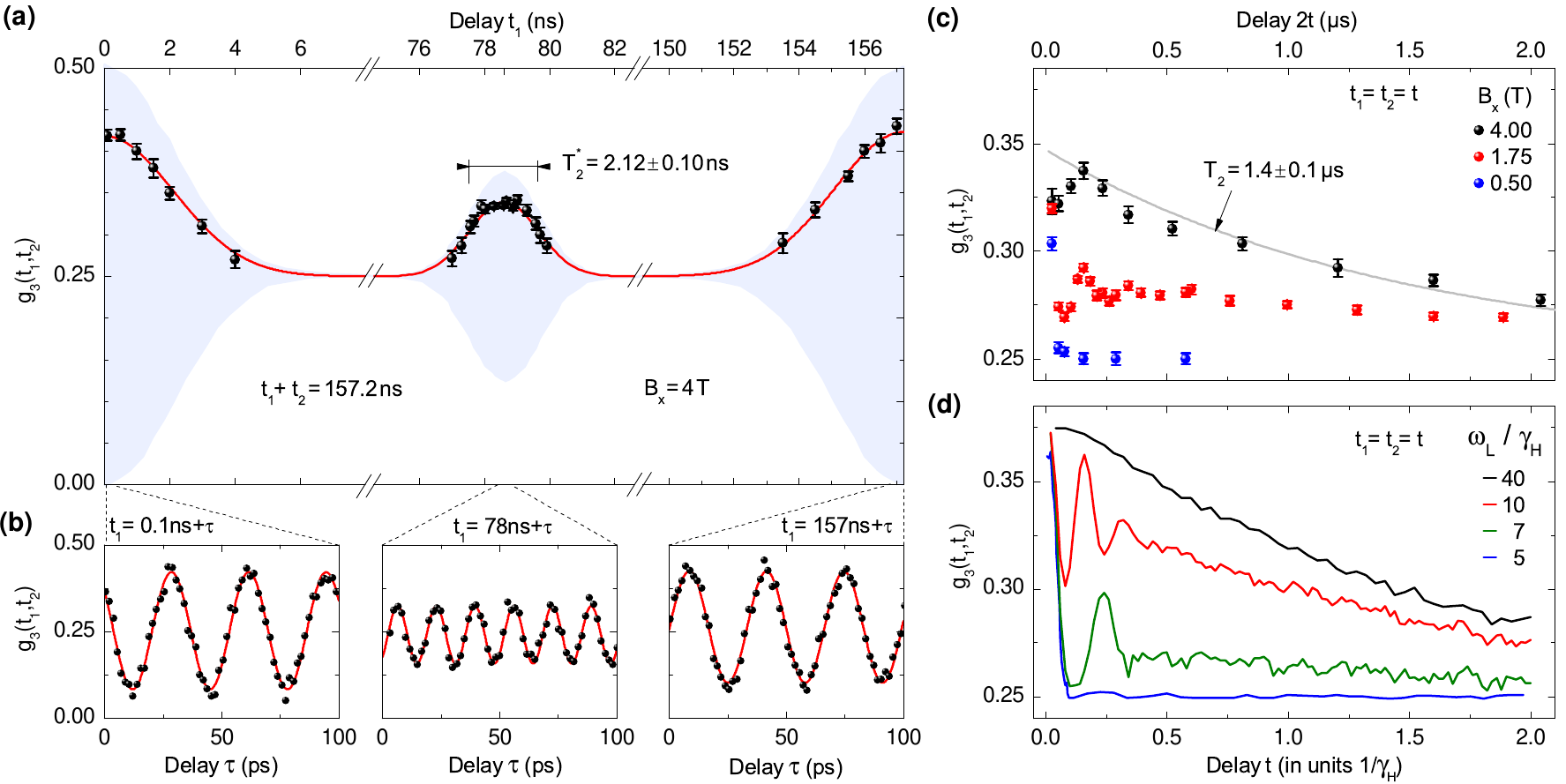}}
\vspace{0.2 cm}
\caption{\label{fig3}
\textbf{Long time behavior of the $g_3$ correlator.}
\textbf{(a)} Experimental data (envelope of Larmor-oscillations) for $g_3$ along the direction $t_1+t_2=157.2~\nano\second$ (anti-diagonal) at $B_x=4~\tesla$. The shaded area indicates the envelope of $g_3$ oscillations with probe fidelities equal to unity.  \textbf{(b)} Oscillations of $g_3$ for $t_1$ at short, intermediate and long time sections, respectively. The data points in (a) are obtained by analyzing the oscillation amplitude using a sinusoidal fit (red line) at different time sections of $t_1$. \textbf{(c)} Experimental data of $g_3(t,t)$ along $t_1=t_2$ direction for different magnetic fields $B_x$. \textbf{(d)} Numerically obtained $g_3(t,t)$ at different magnetic fields. The numerical time is in units of inverse average hyperfine coupling $1/\gamma_H$ of a single nuclear spin (see Methods section), and magnetic field is in units of $\omega_L/\gamma_H$. The number of simulated nuclear spins is $N=900$.}
\end{figure*}

At longer timescales $T_2^*<t_{1,2}<T_2$, i.e. at hundreds of nanoseconds, the oscillation amplitude of $g_3$ along the anti-diagonal direction reflects the dephasing time $T_2^*$. In order to demonstrate this experimentally, we keep the total time $t_1+t_2=157.2~\nano\second$ fixed and tune only the time delay $t_1$. The result of analyzing the oscillation amplitude of $g_3$ along such an anti-diagonal line is shown in Fig.~\ref{fig3}(a) at $B_x=4~\tesla$. Notably, the oscillation amplitude at time instants $t_1 \simeq t_2$ have non-vanishing components for $t_{1,2} \gg T_2^*$ and, hence, are different from classical values according to $g_2(t_1) g_2(t_2)=1/4$, which reflects the quantum nature of the correlator $g_3$. From the width of the Gaussian-like envelope we extract $T_2^*=2.12\pm 0.10~\nano\second$, in perfect agreement with previous $T_2^*$ measurements shown in Fig.~\ref{fig2}(a) and also in Ref.~\cite{Alex}, where we used the same sample but a different measurement method. Fig.~\ref{fig3}(b) shows the experimental data for $g_3$ from which the data points in Fig.~\ref{fig3}(a) are obtained by analyzing the oscillation amplitude. Note the doubled oscillation frequency at times $t_1 \simeq t_2$ in our experiment as predicted theoretically.

In addition to the experimentally measured amplitude of $g_3$ along the anti-diagonal direction, the amplitude along the diagonal direction ($t_1=t_2=t$) as a function of the total time $2t$ is presented in Fig.~\ref{fig3}(c) for different magnetic fields. At high magnetic fields ($B_x =4~\tesla$) the correlator $g_3(t,t)$ decays exponentially with $T_2=1.4 \pm 0.1~\micro\second$. The $T_2$ time here is comparable to previous spin-echo measurements, reported in Ref.~\cite{Alex} ($T_2=1.3~\micro\second$). From Fig.~\ref{fig3}(c) it can be seen that by reducing the magnetic field to $B_x=1.75~\tesla$, $g_3(t,t)$ shows an oscillatory behavior in addition to the $T_2$ decoherence, while  at $B_x=0.5~\tesla$ a quick decay takes place towards the classical limit of $1/4$ after $\sim 40~\nano\second$. Such a reduction of the coherence time with the magnetic field has been observed within the framework of classical spin-echo pulse sequences \cite{Alex}. 

The results of our numerical calculations of $g_3(t,t)$ are presented in Fig.~\ref{fig3}(d).  We simulated the central spin model with the ``Dynamic Mean Field Algorithm" \cite{Al-Hassanieh2006}, which includes hyperfine and quadrupolar couplings of nuclear spins, as in Ref.~\cite{Alex}. We rigorously took quantum measurement  into account (see Methods section). Large timescale separations limited our simulations by $N=900$ nuclear spins vs $N\sim 10^5$ in a real QD. However, the results in Fig.~\ref{fig3}(d) do show qualitatively similar behavior to the experimentally observed data. This confirms that the oscillations of $g_3(t,t)$ at magnetic fields below $4~\tesla$ arise from the combined effect of hyperfine and quadrupole interactions in the nuclear spin bath, in agreement with prior studies \cite{Alex}.

\begin{figure}[t]
\centerline{\includegraphics[width=.6\textwidth]{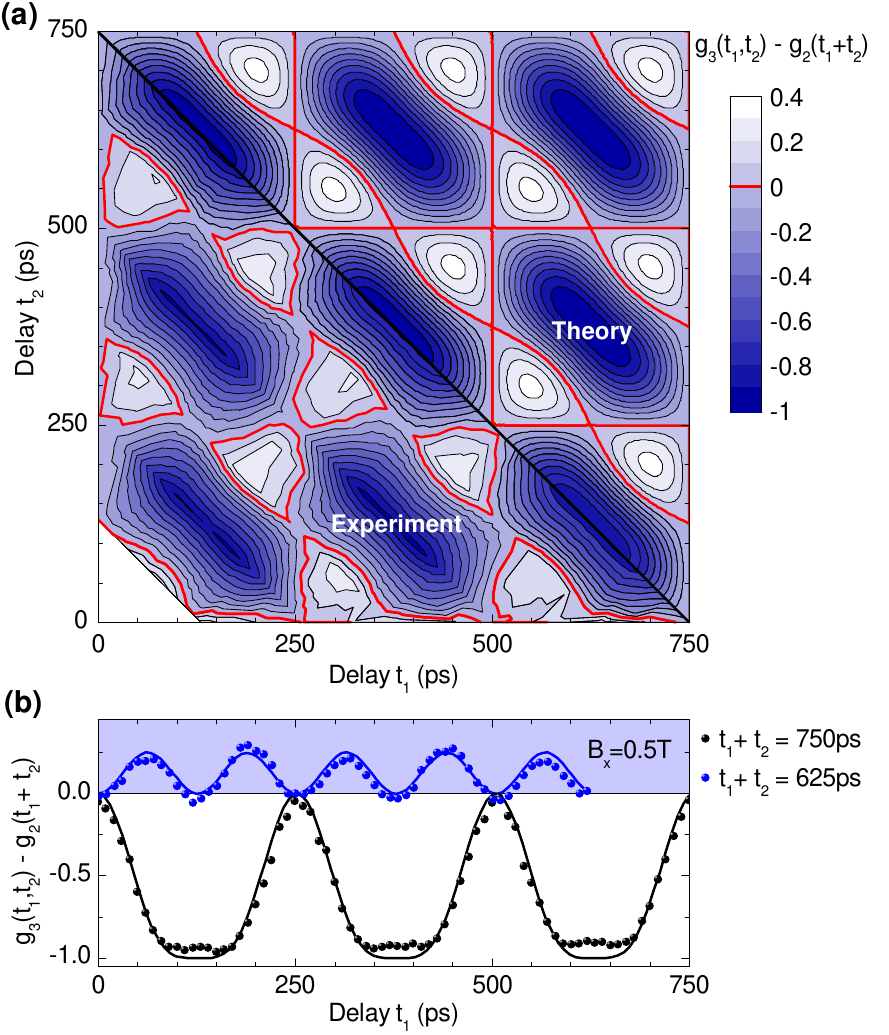}}
\vspace{0.2 cm}
\caption{\label{fig4}
\textbf{Violation of the Leggett-Garg type inequality.}
\textbf{(a)} Contour plot of $g_3(t_1,t_2)-g_2(t_1+t_2)$ at $B_x=0.5~\tesla$. The area within the red contour correspond to data points which violate Eq.~(\ref{leggett1}) and is in very good agreement with theoretical modeling. \textbf{(b)} Experimental data (dots) and theoretical calculation (solid line) corresponding to line-cuts in (a) for $t_1+t_2=690~\pico\second$ and $750~\pico\second$, respectively. Positive values (blue area) are classically forbidden.
}
\end{figure}

Pure quantum behavior of the correlator $g_3(t_1,t_2)$ can be also revealed if we note that, in classical physics, an application of any extra probe pulse would only reduce the probability for a quantum dot to remain in the $1e$-charge state at the end of the measurement sequence. Indeed, imagine that  the spin is always physically present in one of the states $\ket{\ua}$ or $\ket{\da}$, and there is a hidden variable theory that leads to the existence of a joint probability $p_{\alpha, \beta} (t_2+t_1,t_1)$ of observing the values $\alpha,\beta \in \{0,1 \}$ at time moments $t_1+t_2$ and $t_1$. Then $ \la \hat{Q}_{t_{2}+t_1} \hat{Q}_{t_{1}}  \ra \le \la  \hat{Q}_{t_{1}+t_2}  \ra$ and we arrive at a constraint on the correlators of $Q\in \{0,1\}$:
\be
g_3(t_2,t_1) \le g_2(t_1+t_2).
\label{leggett1}
\ee
This relation is of the same origin as the Leggett-Garg inequalities, which are usually formulated for dichotomous variables taking values in $\{ -1, 1\}$ \cite{Leggett1985}. To show the violation of this fundamental inequality we subtracted $g_2(t_1+t_2)$ from $g_3(t_1,t_2)$ measurements. The result is presented in Fig.~\ref{fig4}(a) for $B_x=0.5~\tesla$, where positive values correspond to a violation of the inequality (\ref{leggett1}) (region within the red line in the contour plot). In Fig.~\ref{fig4}(b), two cuts of the experimental data from (a) are presented for $t_1+t_2=690~\pico\second$ and $750~\pico\second$ (dots) together with the corresponding theoretical calculations that assume coherent spin precession (solid lines). The values corresponding to $g_3(t_1,t_2)-g_2(t_1+t_2)>0$ are classically forbidden and demonstrate that the dynamics of a single electron spin cannot be described by a classical theory with hidden variables.

We demonstrated experimentally that fully optical preparation and readout schemes of the electron spin states localized in a semiconductor quantum dot make it possible to characterize higher than 2nd order spin qubit time-correlators for  a wide range of external magnetic fields. We observed effects incompatible with a classical measurement framework. In addition to providing direct evidence for the quantum nature of electron spins our methods provide a paradigm shifting method for measuring the coherence times of qubits which could be applied to a wealth of quantum systems. Measurements of the third order correlator can be used in practice as an alternative to spin echo or dynamic decoupling approaches to determine the ensemble and intrinsic dephasing times, $T_2^*$ and $T_2$, without using coherent spin control sequences.

\section*{Methods}
\subsection{Sample.}
The sample studied consists of a low density ($<~5~\micro\meter^{-2}$) layer of nominally In$_{0.5}$Ga$_{0.5}$As-GaAs self-assembled QDs incorporated into the $140~\nano\meter$ thick intrinsic region of a n-i-Schottky photodiode structure. An opaque gold contact with micrometer sized apertures was fabricated on top of the device to optically isolate single dots. An asymmetric Al$_{0.3}$Ga$_{0.7}$As tunnel barrier with a thickness of $20~\nano\meter$ was grown immediately below the QDs, preventing electron tunneling after exciton generation.

\subsection{Numerical simulations of the central spin model.} 
Our numerical studies of 2$^\text{nd}$ and 3$^\text{rd}$ order correlators were based on simulations of the central spin model that describes a central spin interacting with a large number of nuclear spins via the hyperfine coupling, while nuclear spins experience additional random quadruple fields from strains in a quantum dot. We use the  Hamiltonian, which justification is discussed in more detail in \cite{Sinitsyn2012}, 
\be
\hat{H} = \bB \cdot \hat{ \bS}+\sum_{i=1}^N \gamma_H^i \hat{\bS }\cdot \hat{\bs}^i+\gamma_Q^i (\hat{\bs}^i\cdot \bn^i)+\bb \cdot \hat{\bs}^i,
\label{eff-Ham}
\ee
where $\bB=g_e\mu_e \bB_{ex}$ and $\bb=g_N\mu_N \bB_{ex}$ are Zeeman couplings of the external field to, respectively, electron and nuclear spins; $N$ is the number of nuclear spins. In real quantum dots, $N \sim 10^5$. Numerically, we considered $N=900$, which is large enough to captures the quanlitative behavior of correlators. Spin-1/2 operators $\hat{\bS}$  and  $\hat{\bs}_i$ stand for the central spin and for the $i$th nuclear spin, respectively. $\gamma_H^i$ describes the hyperfine coupling between the central spin and the $i$th nuclear spin. We assume that the distribution of $\gamma_H^i$ is Gaussian, with a characteristic size $\sigma_H$.  The quadrupole coupling is mimicked here by  random static magnetic fields  acting on nuclear spins with the vector $\bn^i$ pointing in a random direction, which is different for different nuclear spins. Parameters $\gamma_Q^i$  mimic the quadrupole couplings, taken from another Gaussian distribution with the mean value $\sigma_Q$.  In the numerical calculations, we take $\sigma_Q=2\sigma_H$. 
 
This model has been already applied to explore spin relaxation effects in the same quantum dot \cite{Alex}. The important addition that we used here to calculate the third order correlator $g_3(t_2, t_1)$, is  the application of the  projection postulate to simulate the quantum measurements by the $\hat{Q}$-operators. We set the central spin to be in the down state  at the initial time moment,  and then after time $t_1$ we record the probability, $P_1$, of finding the central spin in the down state  and then reset the central spin density matrix  to be 
$\hat{\rho}(t_1) = \ket{\downarrow} \bra{\downarrow}$. After another time interval of duration $t_2$, we recorded the probability $P_2$ of finding the central spin in the down state. Then $g_3(t_2, t_1)$ is the average of $P_1 P_2$ over different configurations of randomly chosen initial nuclear spin state vectors. For Fig.~3, averaging was performed  over 30'000 records with different initial conditions. 

\section{Correspondence}
Electronic mail address: jonathan.finley@wsi.tum.de,  nsinitsyn@lanl.gov

\section*{Acknowledgments}
We gratefully acknowledge financial support from the DFG via SFB-631, Nanosystems Initiative Munich, the EU via S3 Nano and BaCaTeC. K.M. acknowledges financial support from the Alexander von Humboldt foundation and the ARO (grant W911NF-13-1-0309). Work at LANL was supported by the U.S. Department of Energy, Contract No. DE-AC52-06NA25396, and the LDRD program at LANL.

\section{Supplementary Information: Quantum Fingerprints in Higher Order Correlators of a Spin Qubit}

\subsection{Electron spin storage and readout scheme}

The electron spin qubit studied in this work is confined in a single self-assembled InGaAs QD incorporated in the intrinsic region of a n-i-Schottky photodiode device next to a AlGaAs tunnel barrier. As illustrated in the schematic band diagram in Fig. 1(a) in the main text, such an asymmetric tunnel barrier design facilitates the control of the electron ($\tau_e$) and hole ($\tau_h$) tunneling time by switching the electric field inside the device. Such a control enables different modes of operation: (i) discharging the QD at high electric fields (not shown in the figure), (ii) optical electron spin initialization realized by applying a single optical picosecond polarized laser pulse (Pump), (iii) spin to charge conversion (Probe 1, 2) after a spin storage time $t_{1,2}$, and (iv) charge readout (not shown in the figure) by measuring the photoluminescence yield obtained by cycling the optical $1e \rightarrow X^{-1}_{3/2}$ transition. The interested reader may refer to Ref.~\cite{Alex} for a detailed illustration of the used spin storage and readout method.

During the reset operation the QD is emptied by an application of high electric fields ($F=190~\kilo\volt\centi\meter^{-1}$) for $500~\nano\second$, enabling a fresh start for the electron spin preparation and readout sequence. After the reset operation we reduce the applied electric field to $F= 70~\kilo\volt\centi\meter^{-1}$. In this regime a $5~\pico\second$ duration $\sigma^+$-polarized laser pulse resonantly drives the $cgs \rightarrow X^0$ transition with $1323.8~\milli\electronvolt$ laser energy (indicated with Pump in Fig. 1(a) in the main text), whereupon an exciton is generated and the hole tunnels out of the QD within $\tau_h = 4~\pico\second$. The applied electric field during the charging regime is chosen such that the hole escape time is much faster than the timescale for exciton fine structure precession ($\sim 150~\pico\second$) providing a spin-$\ket{\downarrow}$ initialization fidelity of $\ge 98\%$. In contrast to the short hole lifetime, electron tunneling is strongly suppressed by the AlGaAs barrier leading to $\tau_\text{e} \gg 10~\micro\second$. To convert the spin information of the resident electron into a charge occupancy and, with this, to probe the electron spin polarization along the optical axis, a second (third) $\sigma^+$-polarized laser pulse with $5~\pico\second$ duration and a laser energy of $1320.4~\milli\electronvolt$ is applied to resonantly excite the $1e \rightarrow X^{-1}$ transition at $F=70~\kilo\volt\centi\meter^{-1}$. During the application of this second (third) optical pulse, the spin information of the resident electron is mapped into a charge occupancy of the QD. Depending on the spin projection of the initialized electron after $t_{1,2}$, the Pauli spin blockade either allows or inhibits laser light absorption. Thus, for electron spin-$\ket{\downarrow}$ projection the QD is charged with $1e$, whereas for spin-$\ket{\uparrow}$ the Pauli spin blockade is lifted, $X^{-1}$ creation is possible and rapid hole tunneling leaves the QD behind charged with $2e$. Finally, the device is biased into the charge readout mode ($F=13~\kilo\volt\centi\meter^{-1}$), where a $1~\micro\second$ duration optical pulse with a laser energy of $1350.6~\milli\electronvolt$ resonantly drives an excited state of the hot trion transition ($1e \rightarrow X_{3/2}^{-1,*}$), probing the charge occupancy of the QD and, therefore, the electron spin polarization after $t_{1,2}$ by measuring the photoluminescense yield from the $X_{3/2}^{-1} \rightarrow 1e$ optical recombination. 

\subsection{Origin of non-classical behavior of  $g_3(t_1,t_2)$}

To understand the non-classical behavior of $g_3(t_1,t_2)$ it is instructive  to relate this correlator to spin correlators at the thermodynamic equilibrium: 
$$
C_2(t)\equiv \la \hat{s}_z(t) \hat{s}_z(0) \ra \equiv {\rm Tr} \Big[ \hat{s}_z \hat{G}(t) \hat{s}_z \Big]/2,
$$
 where $\hat{G}(t)$ was defined in the main text, and we assumed that the equilibrium density matrix is $\hat{\rho}_{\rm eq}=\frac{{1}}{2} \hat{1}$ corresponding to a fully thermal population of the two spin states. The operator $\hat{Q}$ can be expressed in terms of the spin operator $\hat{s}_z$, as $\hat{Q}=\frac{\hat{1}}{2}-\hat{s}_z$. At the thermodynamic equilibrium, the averages of the odd power products of spin operators are practically zero due to the approximate time-reversal symmetry at magnetic field values much smaller than $k_B T$. Disregarding such odd power terms, we find
 \be
%\nonumber 
g_3(t_1,t_2) \approx C_2(t_1) + C_2(t_2) +\frac{1}{4}
+\frac{1}{2} {\rm Tr} \Big[ \hat{s}_z \hat{G}(t_2) \hat{G}(t_1) \hat{s}_z \Big],
\label{third3}
\ee
where    we used the fact that the matrix $\hat{G}$ is doubly stochastic and the equilibrium distribution is invariant under the evolution: $\hat{G}(t)\hat{\rho}_{\rm eq} = \hat{\rho}_{\rm eq}$.

Naively, one can think that since $\hat{Q}$ is linear in the spin operator, and since odd order spin correlators are negligibly small at equilibrium, the correlator $g_3(t_1,t_2)$ should be expressible via the 2$^\text{nd}$ order spin correlators plus a constant. Equation~(\ref{third3}) shows that this is not generally true. For classical nondestructive measurements, the evolution of state probabilities satisfies the convolution rule: $\hat{G}(t_1+t_2)=\hat{G}(t_2) \hat{G}(t_1) $. Hence, classically, the last term in Eq.~(\ref{third3}) would be equal to $C_2(t_1+t_2)$, as expected. However, in quantum mechanics, the convolution rule is valid only for the evolution operators $\hat{U}(t)$ of the Schr\"odinger equation: $\hat{U}(t_1+t_2) = \hat{U}(t_2) \hat{U}(t_1)$, which does not guarantee that such a relation is valid for  probabilities $G_{\alpha \beta}(t)=|U_{\alpha \beta}(t)|^2$.  Thus, we conclude that the deviation of $g_3(t_1,t_2)$ from its classical expression via 2nd order spin correlators at equilibrium corresponds to a purely quantum mechanical measurement effect.

\end{document}